\documentclass[sigconf,natbib=false]{acmart}
\AtBeginDocument{%
  }

\RequirePackage[
  datamodel=acmdatamodel,
  style=acmnumeric,
  sortcites=true
  ]{biblatex}

\usepackage{xcolor}
\usepackage{hyperref}
\usepackage{verbatim}

\begin{document}

%\title{K-12 students' risk perceptions for using artificial
\title{Artificial Intelligence Competence of K-12 Students Shapes Their AI Risk Perception: A Co-occurrence Network Analysis
}
% Please make sure that the short title does not exceed the width of one column
\renewcommand{\shorttitle}{AI Risk Perception}

\author{Ville Heilala}
\orcid{0000-0003-2068-2777}
\affiliation{%
  \institution{University of Jyväskylä}
  \city{Jyväskylä}
  \country{Finland}
}

\author{Pieta Sikstr{\"o}m}
\orcid{0000-0002-2055-7995}
\affiliation{%
  \institution{University of Jyväskylä}
  \city{Jyväskylä}
  \country{Finland}
}

\author{Mika Set{\"a}l{\"a}}
\orcid{0009-0006-7195-9587}
\affiliation{%
  \institution{University of Jyväskylä}
  \city{Jyväskylä}
  \country{Finland}
}

\author{Tommi K{\"a}rkk{\"a}inen}
\orcid{0000-0003-0327-1167}
\affiliation{%
  \institution{University of Jyväskylä}
  \city{Jyväskylä}
  \country{Finland}
}

%This command displays author info in page headers
% Please use the following convention:
% One author: J. Smith
% Two authors: J. Smith and I. Jones
% Three and more authors: J. Smith et al.
\renewcommand{\shortauthors}{V. Heilala et al.}

\begin{abstract}
As artificial intelligence (AI) becomes increasingly integrated into education, understanding how students perceive its risks is essential for supporting responsible and effective adoption. This research aimed to examine the relationships between perceived AI competence and risks among Finnish K-12 upper secondary students (n = 163) by utilizing a co-occurrence analysis. Students reported their self-perceived AI competence and concerns related to AI across systemic, institutional, and personal domains. The findings showed that students with lower competence emphasized personal and learning-related risks, such as reduced creativity, lack of critical thinking, and misuse, whereas higher-competence students focused more on systemic and institutional risks, including bias, inaccuracy, and cheating. These differences suggest that students' self-reported AI competence is related to how they evaluate both the risks and opportunities associated with artificial intelligence in education (AIED). The results of this study highlight the need for educational institutions to incorporate AI literacy into their curricula, provide teacher guidance, and inform policy development to ensure personalized opportunities for utilization and equitable integration of AI into K-12 education.
\end{abstract}




\maketitle

\section{Introduction}

With the rise of artificial intelligence (AI), especially generative AI (GenAI), its influence on society and education is increasingly recognized. However, the integration of artificial intelligence in education (AIED) is not straightforward; on one hand, it has been shown to improve educational outcomes \cite{Tlili2025-hj}, but, on the other hand, there are also risks that can hinder the adoption \parencite{fu2025knowledgeable,Nemorin2023-jq,Topali2025-sl,Colonna2025-nd}. Despite the general view of AI technologies having a positive impact on student learning, their effect particularly on students’ agency and self-regulation is understudied \parencite{darvishi2024impact}. This has raised concerns about how the integration of AI tools affects students' learning experiences, as well as their impact on knowledge development and skill acquisition \parencite{Ukwandu2025-pj,Casal-Otero2023-zv}. 

AI is not perceived solely as harmful or beneficial, but simultaneously as both a risk and an opportunity \parencite{schwesig2023using}. However, perceptions and preferences regarding AI-related risks have received limited scholarly attention \cite{wei2025understanding}. Thereby, the relationship between risk perceptions and individuals' willingness to adopt AI-based applications is understudied \cite{schwesig2023using}. Given that AI is affecting education and how teaching will be organized \cite[e.g.,][]{Heilala2025-wg}, it is crucial to explore further the types of risks that might hinder the adoption of AI tools in learning \parencite{sikstrom2024pedagogical}. Research on risk perception \cite{siegrist2020risk}, especially in the context of AI, is essential for designing effective strategies that support informed decision-making \cite{krieger2024systematic} and promote AI literacy \cite{Casal-Otero2023-zv}. Because "risk is relative to the observer'' \cite[][p.~12]{Kaplan1981-ec}, individual characteristics shape which factors are perceived as risks \parencite{wei2025understanding}. Thus, this exploratory study aims to answer two research questions: First, which of the factors do upper secondary K-12 students perceive as risks? And second, how do K-12 students' self-reported AI competence shape their perceptions of potential AI-related risks?

Finland has been one of the early advocates for adopting digital education solutions and practices\footnote{\href{https://web.archive.org/web/20230930225210/https://www.oph.fi/en/exploring-finnish-digital-education}{Finnish National Agency for Education: Exploring Finnish Digital Education}}. The same applies to AIED, as exemplified by the Finnish National Agency for Education, which has aligned new guidelines for the use of AI in teaching and learning in Spring 2025\footnote{\href{https://web.archive.org/web/20250924085706/https://www.oph.fi/en/artificial-intelligence-education-legislation-and-recommendations}{Finnish National Agency for Education: Artificial intelligence in education – legislation and recommendations} }. AI technologies are novel to K-12 education \cite[e.g.,][]{Filiz2025-lu}, and there is a great demand for understanding how they should be integrated in teaching and learning  \parencite{karan2023potential}. Thus, this research targets general upper secondary education students, as K-12 education has received limited attention in AIED research \cite{Akgun2022-kg,Filiz2025-lu,Heilala2025-wg}. The results contribute to understanding the factors that upper secondary school students perceive as AI-related risks and how their self-reported AI-related competence influences these risk perceptions.

\section{Background}

AI technologies continue to reshape digital society and teaching practices, highlighting the need to understand the factors that facilitate or hinder their use. AI systems offer many benefits, but their integration into K–12 education also poses challenges \cite{yoder2020gaining}. In educational settings, research on AI-related risks has gained popularity, and, for example, six major risk areas have been identified through a systematic literature review in K-12 education: $i)$ privacy and autonomy risks, $ii)$ AI biases, $iii)$ accuracy and functional risks, $iv)$ deepfakes and FATE\textemdash fairness, accountability, transparency, and ethics risks, $v)$ social knowledge and skill-building risks, and $vi)$ risks associated with shifting the teacher's role \cite{karan2023potential}. From the student perspective, the issues may raise several questions, for example: What happens to my sensitive data? Can I trust AI? How capable and accessible is AI? Where and how can I use AI? Am I allowed and able to use AI? Why do we still need teachers?

Risks regarding the use of AI occur at multiple levels. It entails short- and long-term risks, including technology-related risks (e.g., privacy breaches, cyber intrusions, and the inability to control malicious AI) \cite{Zhu2025-wq,Li2023-lq}, educational risks \cite{Zhu2025-wq,Li2023-lq}, economic and societal risks (e.g., job displacement) \cite{Ghotbi2022-tn,Zhu2025-wq}, and ethical risks (e.g., insufficient values and regulations) \parencite{schwesig2023using,Zhu2025-wq,Colonna2025-nd}. Specifically, \textit{systemic risks} refer to broad issues tied to the trustworthiness and functioning of AI itself \cite[e.g.,][]{ai_act,Amoozadeh2024-ah,karan2023potential}, such as bias, inaccuracy, FATE issues \cite{Memarian2023-ke,Akgun2022-kg,Bissessar2023-df,Zhu2025-wq,Li2023-lq,Halton2025-jx,Filiz2025-lu,karan2023potential}, or resource consumption \cite[e.g.,][]{Galaz2021-nh}. \textit{Institutional risks}, in turn, pertain to rules, policies, and fairness within the educational system, including cheating with the aid of AI \parencite{Cavazos2025-zh}, the unequal advantages of using AI (e.g., AI divide and digital divide) \cite{Hammerschmidt2025-ho,Bissessar2023-df,Zhu2025-wq}, or inconsistent guidelines from teachers and schools regarding the use of AI \cite{Corbin2025-xy}. The use of AI also involves \textit{personal risks} that capture individual-level concerns, such as balancing between productivity and misconduct \cite{Corbin2025-xy,Zhu2025-wq,Li2023-lq,Halton2025-jx}, dependence on AI (e.g., metacognitive laziness \cite{Fan2025-cg}), addiction \parencite{Al-Obaydi2025-bz}, as well as potential negative effects on learning, creativity, critical thinking, and personal privacy \parencite{Bissessar2023-df,Zhu2025-wq,Filiz2025-lu,karan2023potential}.

People's perceptions of AI-related risks and opportunities are positively associated with their likelihood of adopting AI \parencite{schwesig2023using}. Risk perception refers to how individuals process information or direct observations about potential hazards and risks of a technology and form judgments and beliefs about its seriousness, likelihood, and acceptability \cite{Renn2013-ff}. Anticipated risks and concerns of AI use can be considered as negative outcome expectancies. An outcome expectancy "is defined as a person's estimate that a given behavior will lead to certain outcomes'' \cite[][p.~193]{Bandura1977-oq}. Negative outcome expectancies (i.e., perceived or experienced risks associated with using AI) can lead to reduced acceptance of AI technologies \cite[e.g.,][]{Yue2023-ib}, and initial negative experiences may be difficult to overcome \cite[e.g.,][]{Henry1995-vt}. Beyond just identifying what people are concerned about and why, research on risk perceptions informs the likelihood of encountering these risks, risk communication, and risk management \parencite{siegrist2020risk}. Furthermore, awareness of potential risks is particularly important in educational settings, because without a clear understanding of AI's roles and functions, educators struggle to implement AI effectively in learning activities \parencite{hwang2020vision}.

\section{Materials and methods}

\subsection{Data collection and participants}

The research was conducted in a Finnish general upper secondary school (ISCED level 3) with approximately 400 students. The school has been engaged in a two-year development project on the educational use of GenAI, funded by the Finnish National Agency for Education. As part of the project, students were introduced to AI in academic counseling sessions, and the school has recently revised its code of conduct to explicitly address the use of AI during studies. 

The questionnaire was administered during classes, and students participated voluntarily, with the option to withdraw at any time without any consequences. An information sheet and a consent form were sent to them two weeks in advance, allowing time to familiarize themselves with the study. In addition, they had the opportunity to ask the researchers for further clarification. A total of 163 students responded to the questionnaire, of whom 47\% identified as women, 51\% as men, and 2\% as non-binary. Median year of birth of the respondents was 2008 (min 2006, max 2008), and approximately half were first-year students (53\% first year, 34\% second year, 13\% third or fourth year).

\subsection{Instruments}

\begin{figure*}[htbp!]
    \centering
    \includegraphics[width=0.95\linewidth]{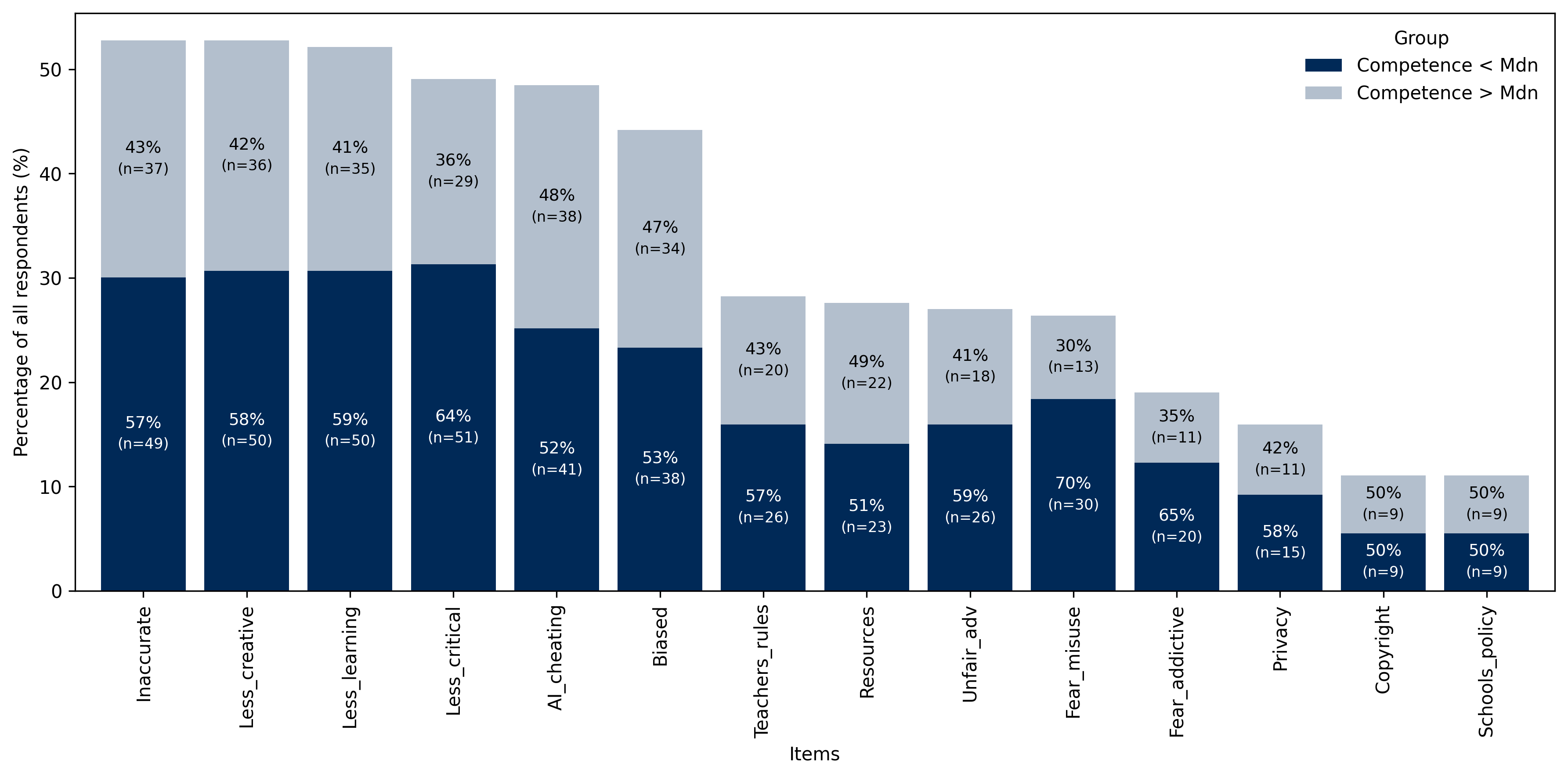}
    \Description{Stacked bar chart showing AI-related concerns segmented by competence group.}
    \caption{Stacked bar chart of AI-related concerns. Bar height indicates the share of all respondents endorsing each item, segmented by competence group. Bar labels show within-item percentages and raw counts by competence group.}
    \label{fig:bar-graph}
\end{figure*}

\subsubsection{AI-related concerns and risks}

The AI-related concerns instrument (Table \ref{tab:ai_concerns}) consisted of 14 binary response items capturing students’ perceptions of potential concerns and risks associated with using AI in schoolwork. Systemic risks included issues related to the functioning of AI, such as bias, inaccuracy, and resource utilization. Institutional risks covered concerns related to rules, policies, and fairness in education, including cheating, unfair advantage, teacher rules, school policies, and copyright. Personal risks reflected individual-level concerns about learning and wellbeing, such as reduced critical thinking, creativity, and learning, as well as fears of misuse, addiction, and privacy violations. Each item was presented as a binary checkbox (0 = not selected / No, 1 = selected / Yes), allowing students to indicate which concerns they considered relevant. Thus, each item can be considered as a binary risk perception variable \cite[e.g.,][]{Lee2015-dw,Lund2014-qo}.

\subsubsection{Artificial intelligence competence}

The AI competence instrument (Table \ref{tab:ai_concerns}) was designed to assess students’ self-reported ability to use AI tools in everyday and educational contexts. It consisted of four items capturing skills in general AI use and in everyday applications \cite[e.g.,][]{Casal-Otero2023-zv}, personalized learning support \cite[e.g.,][]{Wu2025-jq}, and information management \cite[e.g.,][]{Chee2024-uo}, each rated on a 5-point Likert scale. Higher scores on the instrument indicate greater perceived competence in applying AI tools effectively. The confirmatory factor analysis (CFA) utilizing polychoric correlation and robust diagonally weighted least squares (DWLS) estimation \parencite[e.g.,][]{HolgadoTello2010-be,Li2021-xf} supported a unidimensional structure for the four-item AI competence instrument. Model fit ($\chi^2$(2) = 2.56, p = .279; CFI = 1.00; TLI = .999; RMSEA = .041; SRMR = .019) was excellent with respect to conventional criteria \cite[e.g.,][]{Hu1999-do}, indicating that the single-factor model adequately represented the data. The AI competence instrument demonstrated excellent reliability, with the coefficient $\alpha$ indicating a lower-bound reliability estimate $\alpha$ = 0.89, 95\% CI [0.87--0.92].

AI competence sum scores ranged from 4 to 20 (M = 16.8, SD = 2.8, Mdn = 17). For a descriptive overview of AI-related concerns and risks with respect to competence, a median split (Mdn = 17) of AI competence score was used to assign students into low- and high-competence groups, allowing for a comparison of risk perceptions between students with relatively lower and higher self-reported AI competence. Figure \ref{fig:bar-graph} describes the distribution of AI-related concerns among students, showing overall prevalence by binary measurement item and how endorsements were split between low- and high-competence groups. 

\subsection{Co-occurrence network analysis}

Applying a complexity science approach, concerns and risks were analyzed as a complex system \cite[e.g.,][]{Sturmberg2021-tu,Stella2022-he,Borsboom2022-ms}, where potential risks were considered as components of the broader web of beliefs \cite{Porot2021-hj}. A co-occurrence network \cite[e.g.,][]{Bodner2022-mv,Cottica2020-qk} was used to capture the relational structure of the concerns and risks that students hold about AIED. The co-occurrence network graph (Figure \ref{fig:graph-difference}) illustrates how students' co-occurrence of AI-related concerns varies with respect to their AI competence, highlighting patterns in how risks are framed and associated with. The analysis approach draws from Epistemic Network Analysis (ENA) \cite{Shaffer2016-zg} and network psychometrics \cite{Borsboom2022-ms} by modeling co-occurring concerns as interconnected structures \cite[e.g.,][]{Gao2025-jl,Pechey2012-zj}, where edge weights capture conditional associations between items and network metrics quantify the relative importance of concerns within the overall system.

First, binary response items were transformed into a co-occurrence matrix, where each row represented a respondent and each column a pairwise combination of concerns. Cell values were coded as 0 or 1 to indicate whether a co-occurrence was present or absent for the respondent. Before the co-occurrence graph was constructed, a per-respondent $L_2$-normalization \cite[e.g.,][p.~8]{zaki2020data} was applied:
\[
\hat{\mathbf{x}}_i = \frac{\mathbf{x}_i}{\|\mathbf{x}_i\|_2}
= \frac{\mathbf{x}_i}{\sqrt{\sum_{j=1}^d x_{ij}^2}},
\]
where $\mathbf{x}_i$ denotes the vector of concerns selected by student $i$, $x_{ij}$ is the value of item $j$ in that vector, and $d$ is the total number of items. Normalization was applied to control for differences in the number of concerns students selected, ensuring that co-occurrence patterns reflect relative rather than absolute frequencies and preventing overrepresentation of students who endorsed many concerns.

Second, to account for possible confounding influences of gender and year group, regression-adjusted edge weights were estimated by regressing each normalized edge weight on AI competence score while controlling for additional confounding variables using the ordinary least squares (OLS) regression:
\[
w_{ij} = \beta_0 + \beta_1 \,\text{Competence}_i + \beta_2 \,\text{Gender}_i + \beta_3 \,\text{Year\_group}_i + \varepsilon_i,
\]
where $w_{ij}$ is the normalized co-occurrence weight for edge $(i,j)$. The resulting regression coefficient for competence ($\beta_1$) was taken as the undirected edge weight, such that positive values indicate increasing co-occurrence with competence and negative values indicate decreasing co-occurrence. Lastly, to highlight the most salient co-occurrences, edges with absolute values below the global 75th percentile were pruned \cite[e.g.,][]{Cottica2020-qk}, and the remaining network was visualized with node sizes scaled in proportion to the maximum absolute eigenvector centrality across positive and negative subgraphs. Eigenvector centrality \parencite{Bonacich2007-qi,Castro2024-ym} quantifies the relative importance of each node (i.e., concern) within the full co-occurrence network, weighting nodes more highly if they are strongly connected to other highly central concerns.

\section{Results}

\begin{figure*}[htbp!]
    \centering
    \includegraphics[width=0.90\linewidth]{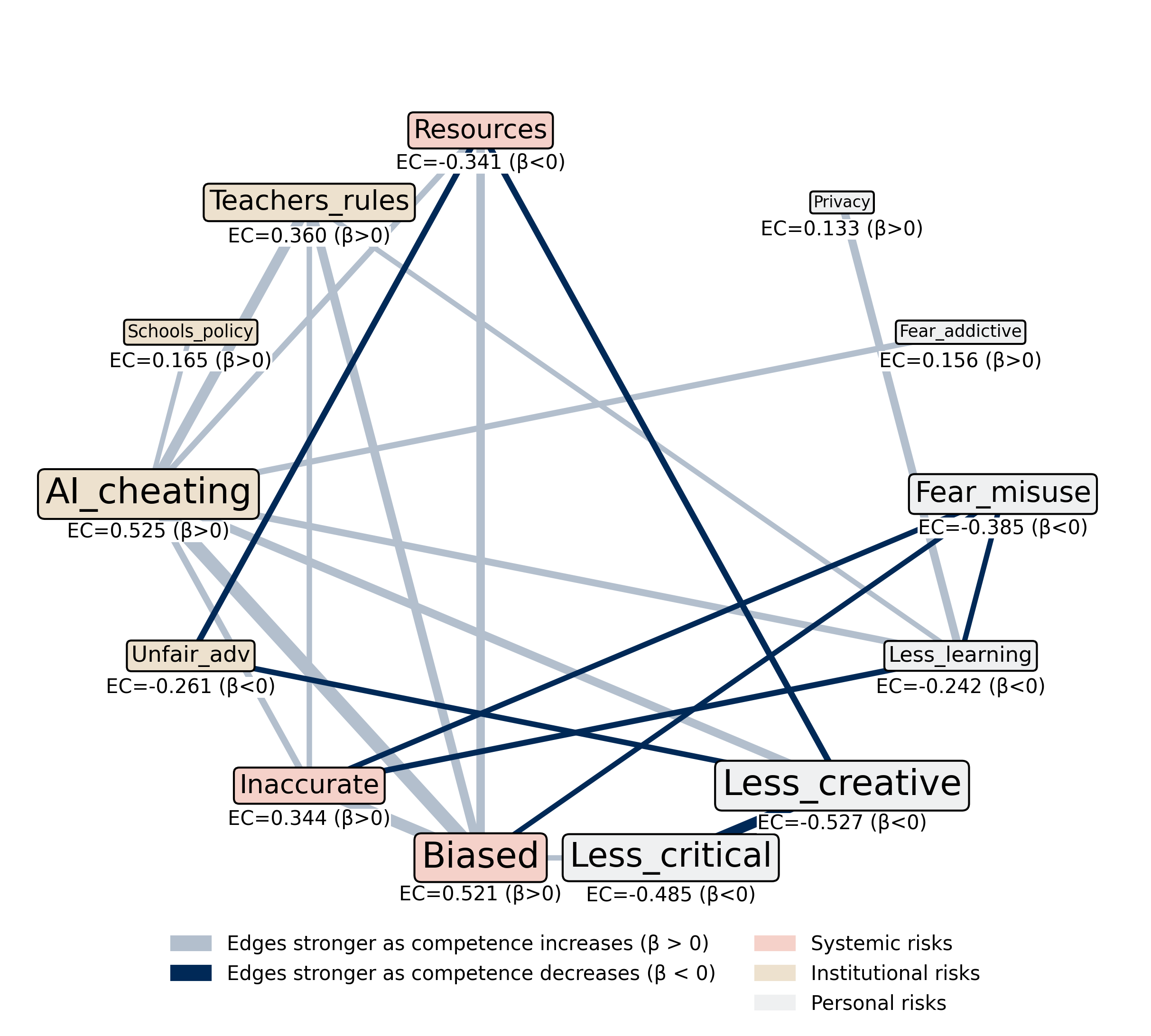}
    \Description{Graph showing co-occurrence network of AI-related concerns.}
    \caption{Co-occurrence network of AI-related concerns.}
    \label{fig:graph-difference}
\end{figure*}

The main concerns receiving the most mentions were AI being inaccurate and biased, AI use for cheating, as well as the risks of learning less and becoming less critical and creative (Figure \ref{fig:bar-graph}). Copyright violations and lack of consistent school rules were the least of the concerns. This pattern suggests that students were primarily concerned with the direct impacts of AI on their own learning quality and fairness, while institutional and legal concerns were perceived generally as less critical.

Students in the low-competence group (Figure \ref{fig:bar-graph}) reported a greater number of concerns across nearly all categories compared to their high-competence peers. They more frequently related AI use to reduced critical thinking, reduced creativity, reduced learning, and fear-based risks, such as misuse and addictive use. Accuracy-related issues were also more common in the low-competence group. Differences in concerns about privacy, copyright, and institutional rules were less pronounced, and they received the fewest mentions overall. In general, the low-competence group of students expressed broader and more numerous concerns, with a strong emphasis on personal learning risks and fear-based issues, while high-competence students reported fewer concerns overall.

The co-occurrence network (Figure \ref{fig:graph-difference}) analysis revealed systematic differences in how students with higher AI competence connected different concerns. Stronger positive associations (edges increasing with competence) were observed (Table \ref{tab:strongest_edges}), particularly between systemic and institutional risks, such as AI\_cheating—Biased, Biased—Inaccurate, and AI\_cheating—Teachers\_rules. Other notable edges included links between bias and fairness-related concerns (Biased—Teachers\_rules, Biased—Resources), as well as connections to personal and learning risks ({AI\_cheating—Less\_creative}, Less\_learning—Privacy). In contrast, several edges decreased in strength with competence. The strongest negative associations were found for Less\_creative—Less\_critical, Less\_creative—Resources, and Less\_creative—Unfair\_adv, suggesting that lower-competence students were more likely to connect creativity-related concerns with fairness and resource issues. Additional decreases were observed in connections involving Fear\_misuse (e.g., with Inaccurate and Less\_learning. Overall, these results indicate that higher-competence students tended to emphasize associations between systemic/institutional concerns (e.g., bias, inaccuracy, and school rules), whereas lower-competence students more strongly connected creativity- and misuse-related concerns.

Eigenvector centrality (EC) values in Figure \ref{fig:graph-difference} showed clear differences in the relative importance of concerns depending on AI competence. In this context, the sign of the eigenvector centrality indicates whether a concern is more structurally central in the co-occurrence network of higher-competence students (positive values) or lower-competence students (negative values). The most central concerns in the high-competence network were AI\_cheating (EC = 0.53) and Biased (EC = 0.52), followed by Teachers\_rules (EC = 0.36) and Inaccurate (EC = 0.34), suggesting that students with higher competence perceive systemic and institutional issues as tightly interconnected. In contrast, the most central concerns in the low-competence network were Less\_creative (EC = -0.53) and Less\_critical (EC = -0.49), indicating that lower-competence students primarily frame risks around threats to their own learning and cognitive development. Other concerns, such as Fear\_misuse (EC = -0.38), Resources (EC = -0.34), and Unfair\_adv (EC = -0.26), also appeared more central in the low-competence group, pointing to a broader emphasis on personal vulnerability and fairness.

\begin{table}[htbp!]
\centering
\caption{Strongest regression-adjusted edges, ranked by $|\beta|$. Direction indicates whether the association increased with AI competence (High $>$ Low) or decreased (Low $>$ High).}
\begin{tabular}{lll}
    \hline
    \textbf{Edge} & \textbf{Direction} & \textbf{$\beta$} \\
    \hline
    AI\_cheating --- Biased           & High $>$ Low  &  0.022 \\
    Biased --- Inaccurate             & High $>$ Low  &  0.017 \\
    AI\_cheating --- Teachers\_rules  & High $>$ Low  &  0.015 \\
    Less\_creative --- Less\_critical & Low $>$ High  & -0.015 \\
    Biased --- Teachers\_rules        & High $>$ Low  &  0.012 \\
    Biased --- Resources              & High $>$ Low  &  0.012 \\
    AI\_cheating --- Less\_creative   & High $>$ Low  &  0.012 \\
    Less\_learning --- Privacy        & High $>$ Low  &  0.011 \\
    AI\_cheating --- Less\_learning   & High $>$ Low  &  0.009 \\
    AI\_cheating --- Inaccurate       & High $>$ Low  &  0.008 \\
    AI\_cheating --- Resources        & High $>$ Low  &  0.008 \\
    AI\_cheating --- Fear\_addictive  & High $>$ Low  &  0.008 \\
    Less\_creative --- Resources      & Low $>$ High  & -0.008 \\
    Less\_creative --- Unfair\_adv    & Low $>$ High  & -0.007 \\
    Inaccurate --- Less\_learning     & Low $>$ High  & -0.007 \\
    Resources --- Unfair\_adv         & Low $>$ High  & -0.007 \\
    Biased --- Fear\_misuse           & Low $>$ High  & -0.006 \\
    Fear\_misuse --- Less\_learning   & Low $>$ High  & -0.006 \\
    Fear\_misuse --- Inaccurate       & Low $>$ High  & -0.006 \\
    Inaccurate --- Teachers\_rules    & High $>$ Low  &  0.006 \\
    Biased --- Less\_critical         & High $>$ Low  &  0.006 \\
    Less\_learning --- Teachers\_rules& High $>$ Low  &  0.006 \\
    AI\_cheating --- Schools\_policy  & High $>$ Low  &  0.006 \\
    \hline
\end{tabular}
\label{tab:strongest_edges}
\end{table}

\section{Discussion}

This study aimed to gain a deeper understanding of the perceptions of AI-related risks that raise concerns among Finnish upper secondary students. The study sheds light on which of the factors students perceive as main AI-related risks, and how students' self-assessed AI competence influences their perception of these risks. The implications of these exploratory results are discussed, and directions for future research are outlined.

The findings suggest that students primarily frame AI-related risks in terms of trust, personal learning, and fairness, while placing less emphasis on broader institutional or legal dimensions. Students were particularly concerned about two \textit{systemic risks}, AI's inaccuracy and bias, and \textit{personal risks} related to learning, specifically the reduction of critical thinking and creativity. From \textit{institutional risks}, students were most concerned about the use of AI for cheating. Taken together, the findings suggest that students are aware of AI's limitations and recognize the potential negative impact on learning. The results of this study further indicate that students' perceptions of risk were linked to their perceived AI competence. \textit{Higher-competence} students primarily emphasize systemic and institutional risks, while \textit{lower-competence} students are more concerned with personal and learning-related risks. Furthermore, lower-competence students reported more concerns spanning nearly all presented risk types. 

However, AI-related risks concern multiple stakeholders. For example, teachers often worry about unintended consequences and the potential reduction of human interaction \cite{Alwaqdani2024-yc,Halton2025-jx,karan2023potential}, and policymakers are concerned about securing an AI-ready workforce and the adoption of technology \cite{Schiff2022-bl}. Thus, AI-related risks should be considered at different levels (i.e., classroom, institutional, and policy-making) to ensure a comprehensive approach to AI adoption in education.

At the classroom level, teachers play a central role in mediating how students engage with AI tools \cite[e.g.,][]{Filiz2025-lu}. \textit{Lower-competence} students were more strongly connected to fears of misuse and addiction with reduced learning, creativity, and critical thinking. Teachers could address these concerns by scaffolding AI use, creating safe spaces for experimentation, and openly discussing both the risks and opportunities associated with AI. Students with \textit{higher-competence} have stronger links between concerns about cheating, bias, teacher rules, and systemic issues such as accuracy and resources. These students may benefit from clear and transparent guidelines on the acceptable use of AI, alongside the development of students' evaluative skills for critically assessing AI outputs. In practice, teachers need to move beyond simply allowing or prohibiting AI use and instead actively integrate it to enhance human creativity and reasoning \cite[e.g.,][]{Brynjolfsson2022-qc}. Self-efficacy beliefs are suggested to mediate AI risk awareness \cite{Wu2025-jq}, and thus, these pedagogical practices may help to mitigate negative outcome expectancies while strengthening students' capabilities for the responsible use of AI.

At the institutional level, the findings highlight the need to explicitly incorporate AI literacy into the curriculum to support the effective implementation of AIED. This study showed that students' perceptions of their AI-related skills shape how they perceive AI risks, aligning with prior research that has established a link between AI self-efficacy and risk awareness \cite{Wu2025-jq}. Perception of risk is dependent on knowledge \cite{Kaplan1981-ec}, and thereby, without systematic and equitable teaching of AI literacy, disparities in knowledge and skills may widen, leading to unequal opportunities to benefit from AI tools (i.e., AI divide) \cite[e.g.,][]{Zhu2025-wq,Hammerschmidt2025-ho}. This echoes earlier debates surrounding the introduction of the internet and search engines into schools: initial fears of plagiarism and misinformation eventually led to the integration of digital and information literacy into curricula \cite{Walraven2009-rd,Ma2007-vn}. A similar transition is now needed for AI, where AI literacy should be integrated into curricula \cite{Casal-Otero2023-zv} by considering students' moral development \cite{Chee2024-uo}, and treated as a transversal skill \cite[e.g.,][]{Bauer2025-ix,Rebelo2025-en}.

At the policy level, policymakers should advance curriculum reform and teacher agency while promoting AI and data literacies \cite{Ifenthaler2024-nr} and ethical practices \cite{Schiff2022-bl} to ensure equitable and effective adoption of AIED \cite{Chee2024-uo}. The EU Artificial Intelligence Act (Regulation (EU) 2024/1689) \cite{ai_act} stresses the importance of ensuring sufficient AI literacy among stakeholders, enabling them to make informed decisions about AI systems \cite[e.g.,][]{Colonna2025-nd}. From a future perspective, failing to institutionalize AI literacy and competence building may leave education systems unprepared for future technological shifts. Research-based pedagogical interventions and guided use of AI in various educational contexts are therefore essential for developing the competence that enables safe, critical, and creative application of AI.

\subsection{Limitations and future research}

This research has certain limitations that should be acknowledged. First, the data were collected in a single Finnish upper secondary school engaged in an AI development project, which may limit the generalizability of the findings to other educational contexts and countries. Second, students' AI competence was measured through self-reports, which may not accurately reflect their actual skills or AI usage practices. Third, the cross-sectional design prevents causal interpretations regarding the relationship between competence and risk perceptions. Lastly, the risks outlined in this research are not an exhaustive list, and other risk perceptions among the students can exist and emerge.

Future research could investigate how risk perceptions \cite[i.e.,][]{siegrist2020risk} among more diverse stakeholders in education intersect and how these perceptions influence their educational experiences. Particular attention could be given to the roles of AI competence and risk perceptions in relation to student agency \cite[e.g.,][]{Heilala2022-xt,Nemorin2023-jq}, as competence may determine whether students approach AI as a supportive tool or perceive it as a limiting factor to autonomy and learning. Also, examining how students balance risks against potential opportunities \cite[e.g.,][]{Halton2025-jx} can provide insights into the conditions under which AI is adopted constructively.

\section{Conclusion}

This research explored Finnish upper secondary students' perceptions of AI-related risks and examined how these perceptions are shaped by their self-reported AI competence. The findings showed that students with lower competence tend to emphasize personal and learning-related risks, while higher-competence students focused more on systemic and institutional concerns. These differences suggest that AI competence plays a crucial role in how students evaluate both the risks and opportunities associated with AI in education. By highlighting the interplay between competence and risk perceptions, the research emphasizes the importance of promoting AI literacy in educational institutions to enable learners to engage critically, responsibly, and productively with AI tools.

\begin{table*}[htbp]
\centering
\caption{Binary items (Yes/No) for AI-related concerns included in the study, with Finnish item wording, English translation, and abbreviation used in analyses. Instruction: FI ``Mitkä ongelmat liittyvät mielestäsi tekoälyn käyttöön koulutehtävissä?'', EN ``What concerns do you think are related to the use of artificial intelligence in school assignments?''}
\begin{tabular}{p{8cm} p{6cm} p{2cm}}
\hline
\textbf{Finnish} & \textbf{English} & \textbf{Abbreviation} \\
\hline
Opettajat voivat pitää tekoälyn käyttöä huijaamisena. & Teachers may consider the use of AI as cheating. & AI\_cheating \\
Tekoälyn käyttö antaa joillekin oppilaille epäreilun edun, jos muut eivät käytä sitä. & The use of AI gives some students an unfair advantage if others do not use it. & Unfair\_adv \\
Tekoäly tuottaa virheellistä tietoa. & AI produces inaccurate information. & Inaccurate \\
Tekoäly tuottaa vinoutunutta tietoa. & AI produces biased information. & Biased \\
Tekoälyn käyttö voi vähentää omaa kriittistä ajatteluani. & AI use may reduce my critical thinking. & Less\_critical \\
Tekoälyn käyttö voi vähentää luovuuttani. & AI use may reduce my creativity. & Less\_creative \\
Tekoälyn käyttö voi vähentää oppimistani. & AI use may reduce my learning. & Less\_learning \\
Pelkään, että vahingossa syyllistyisin väärinkäytöksiin. & I fear I might accidentally misuse AI. & Fear\_misuse \\
Pelkään, että käyttäminen olisi minulle liian koukuttavaa. & I fear using AI would be too addictive for me. & Fear\_addictive \\
Henkilökohtaisia tietojani voi päätyä vääriin käsiin. & My personal data may fall into the wrong hands. & Privacy \\
Tekoälyn tuottama sisältö rikkoo tekijänoikeuksia. & AI-generated content violates copyright. & Copyright \\
Tekoälyn käyttö kuluttaa luonnonvaroja. & AI use consumes natural resources. & Resources \\
Opettajilla ei ole yhtenäisiä sääntöjä tekoälyn käytöstä opiskelussa. & Teachers do not have consistent rules on the use of AI in studying. & Teachers\_rules \\
Kouluilla ei ole ajantasalla olevia käytäntöjä tekoälyn käytöstä opiskelussa. & Schools do not have up-to-date policies on the use of AI in studying. & Schools\_policy \\
\hline
\end{tabular}
\label{tab:ai_concerns}
\end{table*}

\begin{table*}[htbp]
\centering
\caption{Artificial intelligence competence instrument items rated on a 5-point Likert scale. Instruction: FI ``Arvioi tekoälyyn liittyviä tietojasi ja taitojasi. Kuinka paljon olet eri tai samaa mieltä väitteiden kanssa?'', EN `Rate your own knowledge and skills related to artificial intelligence. How much do you agree or disagree with the statements?''}
\begin{tabular}{p{0.5cm} p{8cm} p{7cm}}
\hline
\textbf{\#} & \textbf{Finnish} & \textbf{English} \\
\hline
1 & Minulla on riittävät taidot käyttää tekoälytyövälineitä. & I have sufficient skills to use AI tools. \\
2 & Osaan käyttää tekoälytyökaluja niin, että ne helpottavat arkeani. & I can use AI tools in ways that make my everyday life easier. \\
3 & Osaan hyödyntää tekoälyä omassa oppimisessani. & I can make use of AI in my own learning. \\
4 & Osaan hyödyntää tekoälytyökaluja tiedon etsimisessä. & I can use AI tools for searching information. \\
\hline
\end{tabular}
\label{tab:ai_competence_items}
\end{table*}

\begin{acks}

Support from the Research Council of Finland, under Grant No. 353325, is gratefully acknowledged.

\end{acks}

\section*{Declaration of Generative AI and AI-assisted technologies in the writing process}

During the preparation of this work, the author(s) used Grammarly and ChatGPT 5 in order to proofread and to provide suggestions to enhance the readability and grammatical clarity of the manuscript. After using this tool/service, the author(s) reviewed and edited the content as needed and take(s) full responsibility for the content of the publication.

\printbibliography

%\appendix

%\section{Instruments}\label{instruments}

\end{document}